\documentclass[9pt,twocolumn,twoside]{opticajnl}

\journal{opticajournal} 

\setboolean{shortarticle}{true}

\title{Dark-state photonic entanglement filters}

\author[1,2,*]{Stefano Longhi}

\affil[1]{Dipartimento di Fisica, Politecnico di Milano, Piazza L. da Vinci 32, I-20133 Milano, Italy}
\affil[2]{IFISC (UIB-CSIC), Instituto de Fisica Interdisciplinar y Sistemas Complejos - Palma de Mallorca, Spain}

\affil[*]{stefano.longhi@polimi.it}

\begin{abstract}
Preserving entanglement in the presence of decoherence remains a major challenge for quantum technologies. Recent proposals have employed photonic filters based on anti-parity-time  symmetry to recover certain entangled states, but these approaches require intricate, symmetry-constrained waveguide architectures and precise bath engineering. In this work, we show that such strict non-Hermitian symmetry constraints are not necessary for entanglement filtering. Instead, we identify post-selection and the emergence of dark states -- arising naturally through destructive interference in simple photonic settings -- as the essential mechanisms. By avoiding the need for special bath engineering or non-Hermitian symmetries, our approach significantly simplifies the design and architecture, enhances universality, and extends applicability beyond previously studied dimer configurations. We demonstrate this concept using minimal waveguide network designs, offering a broadly accessible route to robust entanglement filtering.
\end{abstract}

\setboolean{displaycopyright}{false} 

\begin{document}

\maketitle

{\em Introduction.} 
Entanglement, a cornerstone of quantum technologies \cite{R1}, is notoriously fragile and susceptible to decoherence caused by interactions with the environment \cite{R2,R3}. Several strategies have been developed and experimentally demonstrated to mitigate this vulnerability, including quantum error correction \cite{R4,R5}, the use of decoherence-free subspaces \cite{R6,R7}, and dynamical decoupling \cite{R8,R9}. In quantum photonics \cite{R9b}, entanglement filters -- devices that transmit the entangled component of a quantum state while suppressing unwanted classical admixtures -- can enable the recovery of high-fidelity entangled states from mixed inputs \cite{R10,R11,R12,R13,R14}.
A recent study \cite{R15} proposed an entanglement filter based on anti-parity-time (APT) symmetry, implemented in a carefully engineered, lossless waveguide network. That design achieves entanglement recovery in a two-mode (dimer) system by removing classical components, but it relies on enforcing a specific non-Hermitian symmetry and constructing a tailored photonic environment via an isospectral Lanczos transformation.\\ 
A key open question is whether entanglement filters can be realized without relying on non-Hermitian symmetry constraints or intricate reservoir engineering -- and beyond the limited case of dimers -- thereby enabling a more universal and practically viable approach suitable for simpler architectures. In fact, neither non-Hermitian symmetries nor engineered baths appear to be essential. The core mechanism is the presence of a unique dark state -- also known as a decoherence-free, trapped, or bound state in the continuum states \cite{R16,R17,R18,R19,R20,R21,R22,R23,R24,R25,R26,R27,R28,R29,R30,R31,R31b,R31c}-- within each 
$N$-particle sector of the Hilbert space.Under post-selection, these dark states naturally emerge as long-time attractors that remain immune to decoherence, regardless of symmetry or bath details. This mechanism is broadly applicable, as such states arise generically in open quantum systems \cite{R6,R7,R24,R26,R27,R29,R30,R31b,R31c}, including waveguide QED platforms \cite{R24,R26,R29,R30,R31b,R31c} and networks of coupled optical waveguides \cite{R18,R19,R20,R21,R22}, without the need for exotic reservoir engineering.

In this Letter, entanglement filtering is shown to rely solely on the well-established concepts of decoherence-free subspaces and dark-state protection. This leads to a conceptually simple, robust, and widely applicable strategy that naturally generalizes from dimers to arbitrary $M$-mode network architectures. The resulting schemes are scalable, experimentally accessible, and well suited for implementation in photonic systems -- and potentially in other quantum platforms, such as circuit QED -- where similar dark-state mechanisms are present, without relying on non-Hermitian symmetries or complex reservoir design.

{\em Photonic entanglement filter in an optical dimer.}  
A first example of a photonic entanglement filter based on a remarkably simple design is provided by an optical dimer coupled to a semi-infinite, one-dimensional uniform lattice bath, as schematically illustrated in Fig.1(a). This system significantly simplifies the setup used in Ref. \cite{R15}, owing to the homogeneous structure of the lattice bath. It was previously employed in an experiment demonstrating two-photon dark states \cite{R22}.  The two waveguides 1 and 2 of the optical dimer basically behave as two coupled bosonic modes, which are also indirectly and dissipatively coupled  via a common bath. Assuming that the mode propagation constants is the same for all waveguides, photon dynamics in the dissipative optical dimer can be described by the Lindblad master equation with Liouvillian $\mathcal{L}$ (see e.g. \cite{R24,R27,R30})
\begin{eqnarray}
\frac{d \rho}{dz}  =  \mathcal{L} \rho  = -i(H_{eff} \rho-\rho H_{eff}^{\dag})+ \gamma \sum_{n,m=1}^{2} a_m \rho a_n^{\dag}
\end{eqnarray}
In the above equation, ${a}_n$ is the photon annihilation operator in waveguide $n=1,2$,  $\gamma \simeq 2 \kappa^2/J$ is the loss rate of each individual waveguide of the dimer in the bath [here $\kappa$ and $J \gg \kappa$ are the coupling constants indicated in Fig.1(a)], $H_{eff}=\sum_{n,m=1}^{2} (\mathcal{H}_{eff})_{n,m} a_n^{\dag} a_m$ is the effective non-Hermitian Hamiltonian with matrix
\begin{equation}
\mathcal{H}_{eff}=\left(
\begin{array}{cc}
-i \frac{\gamma}{2}  & -i \frac{\gamma}{2}+ \Delta \\
 -i \frac{\gamma}{2} +\Delta & -i \frac{\gamma}{2}
\end{array}
\right),
\end{equation}
and $\Delta$ is the coupling constant between the waveguides 1 and 2. 
A dark state $ | \psi_d \rangle$ corresponds to an eigenstate of $H_{eff}$ such that $ \mathcal{L} | \psi_d \rangle \langle  \psi_d|=0$ \cite{R23}. The pure state $\rho_d=| \psi_d \rangle \langle \psi_d |$ is an eigenstate of the Liouvillian with zero eigenvalue, i.e. it is immune to dissipation.  It can be readily shown that, after letting
\begin{equation}
b^{\dag}=\frac{1}{\sqrt{2}} \left(  \hat{a}_1^{\dag} - a^{\dag}_2 \right) 
\end{equation}
the $N$-photon state
\begin{equation}
| \psi^{(N)}_d \rangle=\frac{1}{\sqrt{N!}} b^{\dag N} |0 \rangle
\end{equation}
is a dark state, for any $N$. This state is clearly an entangled state in the $a_{1,2}$ basis, and it is precisely the same dark state realized in previous experiments \cite{R15,R22} up to $N=2$. Remarkably, in the $N$-particle sector of Hilbert space, $| \psi^{(N)}_d \rangle$ is the unique dark state.
The existence of such dark states does not necessarily require APT symmetry of the non-Hermitian matrix $\mathcal{H}_{eff}$ nor any special tailoring of the bath density of states. In fact, the eigenvalues of $\mathcal{H}_{eff}$ read $\lambda_1=-i \gamma + \Delta$ and $\lambda_2=-\Delta$. APT symmetry occurs whenever the eigenvalue spectrum is invariant under the transformation $ \lambda \leftrightarrow - \lambda^*$, a condition which is met solely in the very accidental case $\Delta=0$ assumed in \cite{R15}. Most importantly, the special network design based on Lanczos transformation used in Ref.\cite{R15}, which ensures a uniform density of states of the bath, is not a necessary requirement.\\ 
 \begin{figure}[h]
 \centering
   \includegraphics[width=0.45\textwidth]{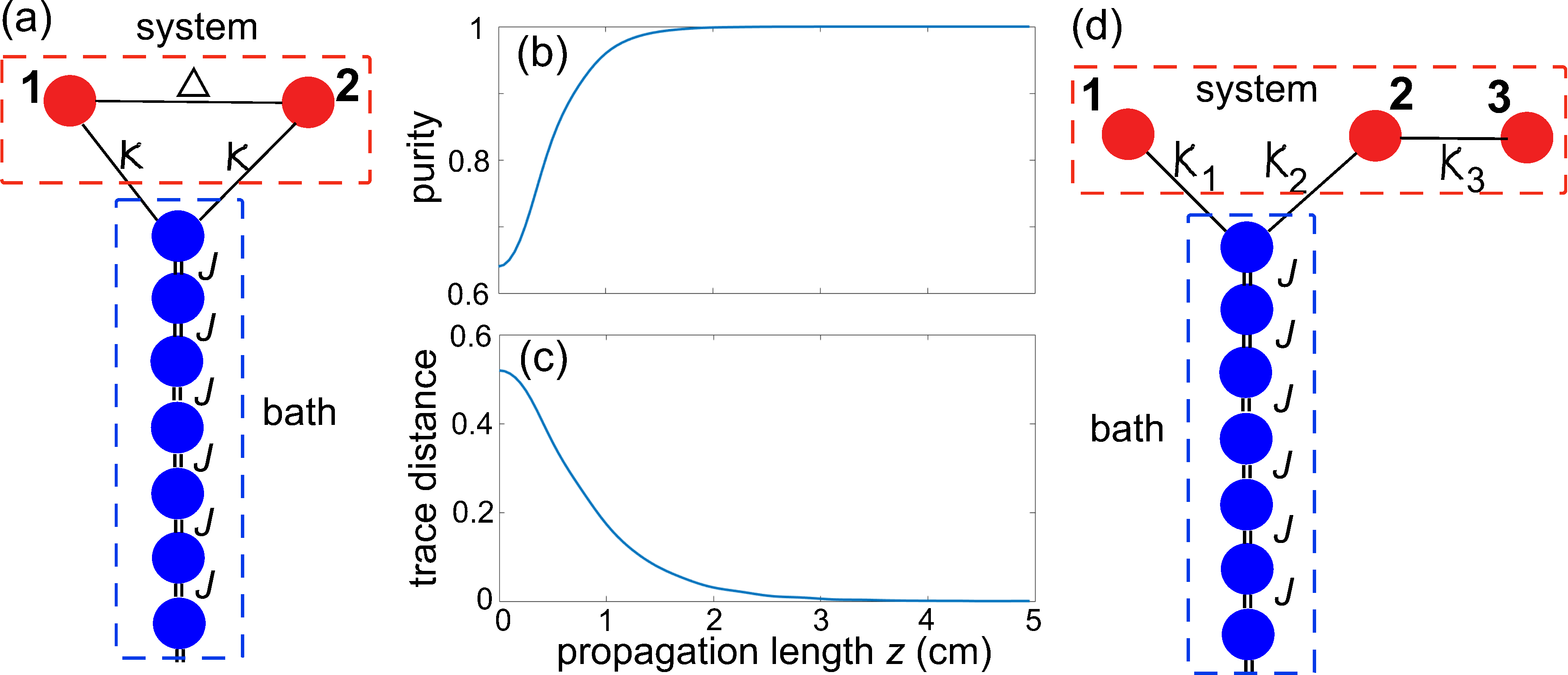}
   \caption{ \small (a) Schematic of a dissipative optical dimer that realizes an entanglement filter after post selection. The system does not display rather generally any non-Hermitian symmetry. The coupling constant $J$ in the waveguide lattice bath is homogeneous. (b,c) Numerically-computed behavior of the purity $\mathcal{P}(z)$ [panel (b)] and trace distance $d(z)$ [panel (c)] of the evolving photon quantum state in the dimer versus propagation length $z$. The input photon state is the mixed state $\rho(z=0)=p | \psi_1 \rangle \langle \psi_1|+(1-p) | \psi_d \rangle \langle \psi_d|$, where $| \psi_d \rangle=(1 / \sqrt{2}) b^{\dag 2} |0 \rangle=(1/2) (|2,0 \rangle+|0,2 \rangle- \sqrt{2} |1,1 \rangle)$ is the two-photon dark state and $|\psi_1 \rangle =|2,0 \rangle$. Parameter values used in the simulations are $J=5.4 \; {\rm cm}^{-1}$, $\kappa=2 \; {\rm cm}^{-1}$, $\Delta=1 \; {\rm cm}^{-1}$ and $p=0.6$. (d) Schematic of a dissipative optical trimer that can sustains a dark state. }
 \end{figure}
To realize entanglement filtering, let us excite at input plane $z=0$ the optical dimer in an arbitrary mixed state belonging to the $N$-particle subspace, i.e. $\rho(z=0)= \sum_{\nu} p_{\nu} | \psi_{\nu} \rangle \langle \psi_{\nu}|$ with $\sum_{n=1,2} \hat{a}^{\dag}_n \hat{a}_n | \psi_{\nu} \rangle=N | \psi_{\nu} \rangle$, and let us consider post selection such that at the output plane $z=z_f$ of the filter we discard the realizations where photon loss occurs, i.e. when a click of photodetectors in the lattice bath waveguides is measured. Under post selection, for a long enough propagation distance, i.e. provided that $z_f$ is sufficiently larger than $ \sim 1 / \gamma$, only the dark state in the $N$-photon space will survive and thus the density matrix $\rho(z_f)$ converges toward the pure state $\rho_d$. 
It should be mentioned that the above analysis describes the photon dynamics for the reduced density matrix of the optical dimer in the Born-Markov approximation using a Lindblad master equation, however the same results can be obtained more generally by considering coherent (Hamiltonian) photon propagation in the full waveguide network using the methods of linear quantum optics (see e.g. \cite{R15,R22,R32a,R32,R33,R34} ). With this method,  the markovian limit $J \gg \kappa$ can be relaxed and the exact evolution dynamics can be computed, as discussed in Sec.1 of the Supplemental document. Also, the existence of the dark state readily follows from the Heisenberg equations of motion of the bosonic creation operators $a_1^{\dag}$ and $a_2^{\dag}$, from which is follows that \cite{R22} $(dc^{\dag}/dz)=0$ with $c^{\dag}=b^{\dag} \exp(i \Delta z)$: this condition ensures that the state $ | \psi_d \rangle$, given by Eq.(4), is a trapped state, effectively decoupled from the bath. An example of entanglement filtering of an initial mixed state  in the $N=2$ particle sector is shown in Figs.1(b) and (c). 
The filtered state  is the dark state $| \psi_d \rangle$  given by Eq.(4) with $N=2$, i.e. $ |\psi_d \rangle=(1/2) | 2,0 \rangle+(1/2) | 0,2 \rangle-(1/ \sqrt{2}) |1,1 \rangle$, where we have set $| n,m \rangle \equiv 
a_1^{\dag n} a_2^{\dag m} / (\sqrt{n!m!}) |0 \rangle$.
 Numerical simulations have been performed beyond the weak-coupling and markovian approximations, considering the Hamiltonian photon propagation in the full waveguide network and calculating the reduced density operator $\rho(z)$ as described in the Supplemental document.
 Figure 1 depicts the numerically-computed behavior of the purity [Fig.1(b)]
\begin{equation}
\mathcal{P}(z)={\rm Tr} \left( \rho^2(z) \right)
\end{equation}
 and the trace distance [Fig.1(c)]
\begin{equation}
d(z)={\rm Tr} \sqrt{\left( \rho(z)-\rho_d \right)^2}
\end{equation}
of the evolving quantum state versus propagation length $z$ under post selection. The values of coupling constants $J$, $\kappa$ and $\Delta$ used in the numerical simulations are typical for waveguide lattices manufactured by the femtosecond-laser-writing technology \cite{R15,R22}. Both trace distance $d(z)$ and purity $\mathcal{P}(z)$ are  bounded in the interval $(0,1)$. The former gives an information on how much the state is mixed, whereas the latter provides a measures of the distance between the quantum states $\rho(z)$ and the target entangled pure state $\rho_d$, with $d(z)=0$ if and only if $\rho(z)=\rho_d$.  The optical dimer is initially excited, at $z=0$, by the mixed state $\rho(z=0)=p | \psi_1 \rangle \langle \psi_1|+(1-p) | \psi_d \rangle  \langle \psi_d|$ with $|\psi_1 \rangle=|2,0 \rangle$ and $p=0.6$. Figures 1(b) and (c) clearly show that, after an initial transient, the post-selected density matrix $\rho(z)$ rapidly converges toward the pure state $\rho_d=| \psi_d \rangle \langle \psi_d|$.\\
\\
\begin{figure*}
 \centering
   \includegraphics[width=0.95\textwidth]{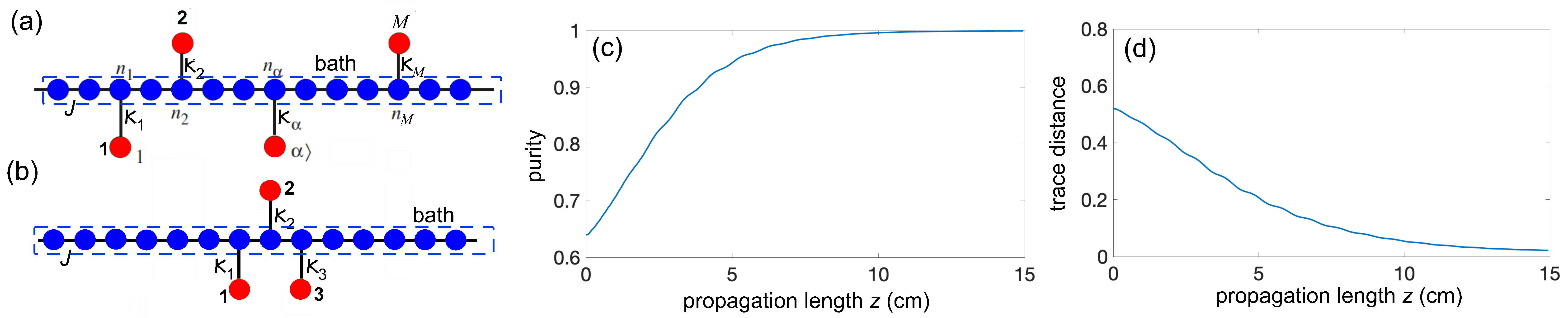}
   \caption{ \small (a) Schematic of a photonic network comprising $M$ waveguides (the system) side-coupled to a one-dimensional tight-binding lattice with uniform coupling constant (the bath). (b) Schematic of the network for $M=3$  and for $n_3-n_2=n_2-n_1=1$,  that realizes a dissipative optical trimer. The system does not display rather generally any non-Hermitian symmetry, however it can sustain a dark state whenever the condition (15) given in the main text is satisfied. (c,d) Numerically-computed behavior of the purity $\mathcal{P}(z)$ [panel (c)] and trace distance $d(z)$ [panel (d)] of the evolving photon quantum state versus propagation length $z$. The input photon state is the mixed state $\rho(z=0)=p | \psi_1 \rangle \langle \psi_1|+(1-p) | \psi_d \rangle \langle \psi_d|$, where $| \psi_d \rangle$ is the one-photon dark state and $|\psi_1 \rangle =|1,0,0 \rangle$. Parameter values used in the simulations are $J=10 \; {\rm cm}^{-1}$, $\kappa_1=\kappa_2=\kappa_3= 2\; {\rm cm}^{-1}$, $\omega_1=\omega_3=\sqrt{2}J$, $\omega_2= \omega_1-\kappa_2^2 / \omega_1$, and $p=0.6$. }
 \end{figure*}
{\em Photonic entanglement filter in an optical trimer}. Several simple photonic architectures that exhibit dark states -- without relying on any non-Hermitian symmetry -- can be constructed by extending the basic setup of Fig.1(a). This provides a universal and more general framework for implementing entanglement filters beyond the dimer case. For instance, the dissipative trimer shown in Fig.1(d) functions effectively as an entanglement filter, even though its underlying non-Hermitian Hamiltonian lacks any symmetry constraints (see Sec.2 of the Supplemental document for technical details). 
A different and more general photonic architecture that can sustain dark states, and thus suited for realizing quantum entanglement filtering under post selection, is provided by a set of $M$ optical waveguides side-coupled to a one-dimensional waveguide lattice \cite{R20} or a slab waveguide \cite{R19}, as schematically shown in Fig.2(a). This configuration is analogous to the typical setting used in waveguide QED to realize dark states, where a set of $M$ quantum emitters (such as superconducting qubits) are indirectly coupled via a common bus waveguide \cite{R24,R26,R27,R29,R30,R31b,R31c,R35}. It can also model indirect coupling of giant atoms \cite{R27}. In such architecture, decoherence-free states can exist without any special non-Hermitian symmetry requirement nor complex bath engineering. Specifically, let us indicate by $J$ the coupling constant between adjacent waveguides in the lattice bath, by $\kappa_{\alpha}$ ($\alpha=1,2,3,...,M$) the coupling constant of the side waveguide $\alpha$ with the lattice and 
by $\omega_{\alpha}$ its propagation constant mismatch from the lattice waveguides ($|\omega_{\alpha}|< 2J$). 
The propagation constant mismatch $\omega_{\alpha}$ is assumed to be of the form $\omega_{\alpha}= \Omega+\Omega_{\alpha}$, where $\Omega$ is  a uniform bias  of order $\sim J$  and $| \Omega_{\alpha}| \ll \Omega$  are small deviations from be bias. The lateral waveguides are coupled to the lattice bath at the sites $n_{1}$, $n_2$,..., $n_M$, as schematically shown in Fig.2(a). In the Born-Markov approximation $\kappa_{\alpha} \ll J$ and neglecting delay effects, the photon field in the lateral waveguides, described by the reduced density operator $\rho(z)$,  propagates according to the master equation (see e.g. \cite{R24,R30,R31c})
\begin{eqnarray}
\frac{d \rho}{dz}  =  \mathcal{L} \rho  = -i(H_{eff} \rho-\rho H_{eff}^{\dag})+ \sum_{\alpha,\beta=1}^{M} \gamma_{\alpha,\beta} a_{\alpha} \rho a_{\beta}^{\dag}
\end{eqnarray}
with an effective non-Hermitian Hamiltonian $H_{eff}=\sum_{\alpha,\beta=1}^M (\mathcal{H}_{eff})_{\alpha,\beta} a^{\dag}_{\alpha} a_{\beta}$ and
\begin{equation}
(\mathcal{H}_{eff})_{\alpha,\beta}= J_{\alpha,\beta}-i \frac{\gamma_{\alpha,\beta}}{2} .
\end{equation}
The explicit expression of the dissipative and coherent terms $\gamma_{\alpha,\beta}$ and $J_{\alpha,\beta}$ entering in the effective non-Hermitian Hamiltonian can be derived using the method detailed in Ref.\cite{R20} and read
\begin{eqnarray}
J_{\alpha,\beta} & = &  \omega_{\alpha} \delta_{\alpha,\beta}+ \rho {\kappa_{\alpha} \kappa_{\beta}} \sin (k_0| n_{\alpha}-n_{\beta}|)  \\
\frac{\gamma_{\alpha,\beta}}{2} & = & \rho \kappa_{\alpha} \kappa_{\beta}  \cos (k_0| n_{\alpha}-n_{\beta}| )
\end{eqnarray}
where we have set
\begin{equation}
\rho=\frac{1}{\sqrt{4J^2-\Omega^2}} \; ,\;\; 
k_0= -\frac{\pi}{2}+ {\rm atan} \left(  \frac{\Omega}{\sqrt{4 J^2-\Omega^2}} \right) 
\end{equation}
The conservative and dissipative terms in the non-Hermitian Hamiltonian, as given by Eqs.(9) and (10), have the usual form found in waveguide QED models for point-like quantum emitters (see e.g. \cite{R30,R31c}). Note that, since $ 2 J \cos k_0=\Omega$, physically $k_0$ is the Bloch wave number of the photon modes excited in the lattice bath in the relaxation process.  A dark state arises whenever the non-Hermitian matrix $\mathcal{H}_{eff}$ has an eigenvalue with vanishing imaginary part, i.e. a non-decaying eigenstate.  The existence of a dark state is not related to any non-Hermitian symmetry of $\mathcal{H}_{eff}$, such as APT symmetry. To illustrate this point, let us consider the case $M=3$ with $n_2-n_1=n_3-n_2=1$, i.e. an optical trimer where three waveguides are indirectly coupled via a common lattice bath; see Fig.2(b). In this case, provided that the conditions 
\begin{equation}
\omega_3= \omega_1 \; , \; \; \omega_2=\omega_1-{\kappa_2^2}/{\omega_1}.
\end{equation} 
are met, the state
\begin{equation}
| \psi_d^{(N)} \rangle= \frac{1}{\sqrt{N!}} b^{\dag N}|0 \rangle
\end{equation}
is a dark state in the $N$-particle sector, effectively decoupled from the bath, where $b^{\dag}$ is the creation operator of the dressed (entangled) state given by
\begin{equation}
b^{\dag}=
\mathcal{N} \left( a_1^{\dag}- \frac{\kappa_1  \omega_1}{J \kappa_2} a_2^{\dag} + \frac{\kappa_1}{\kappa_3} a_3^{\dag}\right)  
\end{equation}
and $\mathcal{N}$ is a normalization factor such that $[b, b^{\dag}]=1$. Technical details are given in the Supplemental document. For example, we can satisfy condition (12) assuming $\kappa_1=\kappa_2=\kappa_3=\kappa$, $\omega_1=\omega_3=\Omega= \sqrt{2}J$, and 
$\omega_2=\Omega- \kappa^2 / \Omega$,
corresponding to $k_0= -\pi /4$. For such a choice of parameters, the dark state (13) with $N=1$ takes the form
\begin{equation}
| \psi_d \rangle=\frac{1}{2} \left(  a_1^{\dag}- \sqrt{2}a_2^{\dag} +a_3^{\dag} \right) |0 \rangle
\end{equation}
and the corresponding non-Hermitian matrix reads
\begin{equation}
H_{eff}= \left(
\begin{array}{ccc}
\Omega-i \frac{\kappa^2}{ \sqrt{2}J} & -i \frac{\kappa^2}{2J}(1-i) & - \frac{\kappa^2}{\sqrt{2}J}  \\
-i \frac{\kappa^2}{2J} (1-i) & \Omega- \frac{\kappa^2}{\sqrt{2}J} (1+i) & -i \frac{\kappa^2}{2J} (1-i) \\
-\frac{\kappa^2}{\sqrt{2} J } &- i \frac{\kappa^2}{2J} (1-i) & \Omega-i \frac{\kappa^2}{\sqrt{2} J } 
\end{array}
\right).
\end{equation}
Such a matrix does not display any APT symmetry, as one can see after inspection of its eigenvalues, given by
\begin{equation}
\lambda_1=\Omega \; ,\;\; \lambda_2=\Omega- \frac{ \sqrt{2}\kappa^2}{J} (1+i) \; ,\; \;  \lambda_2=\Omega+ \frac{\kappa^2}{ \sqrt{2}J} (1-i).
\end{equation}
 The existence of the dark state can be proven, beyond the limit of validity of the Lindblad master equation (7),
 by considering coherent (Hamiltonian) photon propagation in the full waveguide network. In this analysis the dark state can be readily found by decoupling the photon dynamics in a subset of waveguides  from all other waveguides of the network, as discussed in Sec.1 of the Supplemental document.  
To realize entanglement filtering,  we assume that the trimer is excited  at input plane $z=0$  by an arbitrary mixed state belonging to the $N$-photon subspace, i.e. $\rho(z=0)= \sum_{\nu} p_{\nu} | \psi_{\nu} \rangle \langle \psi_{\nu}|$ with $\sum_{n=1,2,3} \hat{a}^{\dag}_n \hat{a}_n | \psi_{\nu} \rangle=N | \psi_{\nu} \rangle$. Under post selection, i.e. discarding the realizations where photon loss occurs, for a long enough propagation distance $z_f$ only the dark state in the $N$-photon space will survive and thus the density matrix $\rho(z_f)$ converges toward the pure state $\rho_d$. An illustrative example of entanglement filtering in the $N=1$ particle sector is shown in Fig.2(c,d). 
The filtered state  is the dark state $| \psi_d \rangle$  given by Eq.(15). Numerical simulations have been performed beyond the weak-coupling limit, considering the Hamiltonian photon propagation in the full waveguide network and calculating the reduced density  operator $\rho(z)$ as described in Sec.1 of the Supplemental document.
Figure 2(c,d) depicts the numerically-computed behavior of the purity [Fig.2(c)] and trace distance [Fig.2(d)] of the evolving quantum state under continuous monitoring and post selection. The input photon state, at $z=0$, is the mixed state $\rho(z=0)=p | \psi_1 \rangle \langle \psi_1|+(1-p) | \psi_d \rangle  \langle \psi_d|$, where $| \psi_d \rangle$ is the one-photon dark state, $|\psi_1 \rangle=a_1^{\dag} |0 \rangle \equiv |1,0,0 \rangle $ and $p=0.6$. The simulations clearly show that, after an initial transient, the post-selected density matrix $\rho(z)$ converges toward the pure state $\rho_d=| \psi_d \rangle \langle \psi_d|$.\\
\\
{\em Conclusion.}
To conclude, it has been shown that high-fidelity entanglement filtering -- previously demonstrated in constrained architectures \cite{R15} -- can be achieved in simple photonic networks without invoking non-Hermitian symmetries or engineered environments. The key mechanism is the existence of a unique dark state, which remains protected under dissipation through post-selection, enabling probabilistic entanglement recovery. This strategy builds on well-established principles such as decoherence-free subspaces and dark-state protection, without requiring system-specific symmetries or tailored reservoirs. As such, it offers a universal and conceptually transparent approach to mitigating decoherence in entangled photonic states. The method relies on minimal architectural complexity -- demonstrated here in dimers and trimers, and generalizable to arbitrary $M$-mode networks -- making it readily implementable across quantum hardware platforms.
In particular, the protocol is well-suited for current integrated photonic technologies, where high-precision waveguide fabrication is routine. Moreover, because dark-state physics arises in other open quantum systems -- such as waveguide QED and circuit QED -- the method is broadly transferable to platforms where complex bath engineering like Lanczos transformations are impractical. Its simplicity, generality, and scalability make it a promising route for robust entanglement distribution in emerging quantum technologies.\\
\\ 
\noindent
\small
{\bf Disclosures}. The author declares no conflicts of interest.\\
{\bf Data availability}. No data were generated or analyzed in the presented research.\\
{\bf Funding}. Agencia Estatal de Investigacion (MDM-2017-0711).\\
{\bf Supplemental document}. See Supplement 1 for supporting content.\\

\newpage


 {\bf References with full titles}\\
 \\
 \noindent
 1. N. Gisin and R. Thew, Quantum communication, Nat. Photon. {\bf 1},
165 (2007).\\
 2. W. H. Zurek, Decoherence, einselection, and the quantum origins of the classical, Rev. Mod. Phys. {\bf 75}, 715 (2003).\\
 3. M.A. Schlosshauer, Decoherence, the measurement problem, and interpretations of quantum mechanics, Rev. Mod. Phys. {\bf 76}, 1267 (2005).\\ 
 4. J. Chiaverini, D. Leibfried, T. Schaetz, M. D. Barrett, R. B. Blakestad, J. Britton, W. M. Itano, J. D. Jost, E. Knill, C. Langer, R. Ozeri, and D. J. Wineland, 
Realization of quantum error correction, Nature {\bf 432}, 602 (2004).\\
 5. M. D. Reed, L. DiCarlo, S. E. Nigg, L. Sun, L. Frunzio, S.M. Girvin, and R. J. Schoelkopf, 
Realization of three-qubit quantum error correction with superconducting circuits,
Nature {\bf 482}, 382 (2012).\\
6. D. A. Lidar, I. L. Chuang, and K. B. Whaley, Decoherence-Free Subspaces for Quantum Computation,
Phys. Rev. Lett. {\bf 81}, 2594 (1998).\\
7. P.G. Kwiat, A.J. Berglund, J.B. Altepeter, and A.G. White, Experimental verification of decoherence-free subspaces,
Science {\bf 290}, 498 (2000).\\
8. L. Viola, E. Knill, and S. Lloyd,  Dynamical Decoupling of Open Quantum Systems, Phys. Rev. Lett. {\bf 82}, 2417 (1999).\\
9. J. Bylander, S. Gustavsson, F. Yan, F. Yoshihara, K. Harrabi, G. Fitch, D.G. Cory, Y. Nakamura, J.-S. Tsai, and W.D. Oliver,  Noise spectroscopy through dynamical decoupling with a superconducting flux qubit. Nature Phys {\bf 7}, 565 (2011).\\
10. J. Wang, F. Sciarrino, A. Laing, and  M.G. Thompson, Integrated photonic quantum technologies, 
Nature Photon. {\bf 14}, 273 (2020).\\
11. H.F. Hofmann and S. Takeuchi, Quantum filter for nonlocal
polarization properties of photonic qubits. Phys. Rev. Lett. {\bf 88},
147901 (2002).\\
12. R. Okamoto, J.L. O'Brien, H. F. Hofmann, T. Nagata, K. Sasaki, and S. Takeuchi, An entanglement filter,
Science {\bf 323}, 483 (2009).\\
13. X.-Q. Zhou, T.C. Ralph, P. Kalasuwan, M. Zhang, A. Peruzzo, B.P. Lanyon, and J.L. O'Brien, Adding control to arbitrary unknown quantum operations,
Nature Commun. {\bf 2}, 413 (2011).\\
14. X. Qiang, X. Zhou, J. Wang, C.M. Wilkes, T. Loke, S. O'Gara, L. Kling, G.D. Marshall, R. Santagati, T.C. Ralph, J.B. Wang, J.L. O'Brien, M.G. Thompson, and J.C.F. Matthews,
Large-scale silicon quantum photonics implementing arbitrary two-qubit processing,
Nature Photon. {\bf 12}, 534 (2018).\\
15. G.-S. Ye, B. Xu, Y. Chang, S. Shi, T. Shi, and L. Li, A photonic entanglement filter with Rydberg atoms, Nature Photon. {\bf 17}, 538 (2023).\\
16. M.A. Selim, M. Ehrhardt, Y. Ding, H.M. Dinani, Q. Zhong, A. Perez Leija, S.K. Ozdemir, M. Heinrich, A. Szameit, D.N. Christodoulides, and M. Khajavikhan, Selective filtering of photonic quantum entanglement via anti-parity-time symmetry, Science {\bf 387}, 1424 (2025).\\
17. P. L. Knight, M.A. Lauder, and B.J. Dalton, Laser-induced continuum structures, Phys. Rep. {\bf 190}, 1 (1990).\\
18. G. Sudarshan, in {\it Field Theory, Quantization and Statistical Physics} edited by E.~Tirapegui (D. Reidel Publishing, 1988), pp. 237-245.\\
19. S. Longhi, Bound states in the continuum in a single-level Fano-Anderson model, Eur. Phys. J. B {\bf 57}, 51 (2007).\\
20. S. Longhi, Optical analog of population trapping in the continuum: Classical and quantum interference effects, Phys. Rev. A {\bf 79}, 023811 (2009).\\
21.S. Longhi, Optical analogue of coherent population trapping via a continuum in optical waveguide arrays, J. Mod. Opt. {\bf 56}, 729 (2009).\\
22.F. Dreisow, A. Szameit, M. Heinrich, R. Keil, S. Nolte, A. T\"unnermann, and S. Longhi, Adiabatic transfer of light via a continuum in optical waveguides, Opt. Lett. {\bf 34}, 2405 (2009).\\
23.A. Crespi, L. Sansoni, G. Della Valle, A. Ciamei, R. Ramponi, F. Sciarrino, P. Mataloni, S. Longhi, and R. Osellame, 
Particle Statistics Affects Quantum Decay and Fano Interference,
 Phys. Rev. Lett. {\bf 114}, 090201 (2015).\\
24.R.I. Karasik, K.-P. Marzlin, B.C. Sanders, and K.B. Whaley,
Criteria for dynamically stable decoherence-free subspaces and incoherently generated
coherences, Phys. Rev. A {\bf 77}, 052301 (2008).\\
25. K. Lalumi\'ere, B.C. Sanders, A.F. van Loo, A. Fedorov, A. Wallraff, and A. Blais, Input-output theory for waveguide QED with an ensemble of inhomogeneous atoms, Phys. Rev. A {\bf 88}, 043806 (2013).\\
26. D.A. Lidar,
Review of Decoherence Free Subspaces, Noiseless Subsystems, and Dynamical Decoupling, Adv. Chem. Phys. {\bf 154}, 295 (2014).\\
27. V Paulisch, H J Kimble and A Gonz\'alez-Tudela,
Universal quantum computation in waveguide QED using decoherence free subspaces,  New J. Phys. {\bf 18}, 043041 (2016).\\
28. A.F. Kockum, G. Johansson, and F. Nori, 
Decoherence-Free Interaction between Giant Atoms in Waveguide Quantum Electrodynamics, Phys. Rev. Lett. {\bf 120} 140404 (2018).\\
29. S.L. Wu, L.C. Wang, and X.X. Yi, 
Time-dependent decoherence-free subspace, J. Phys. A: Math. Theor. {\bf 45}, 405305 (2021).\\
30. M. Zanner, T. Orell, C.M.F. Schneider, R. Albert, S. Oleschko, M.L. Juan, M. Silveri, and G. Kirchmair, 
 Coherent control of a multi-qubit dark state in waveguide quantum electrodynamics, Nat. Phys. {\bf 18}, 538 (2022).	\\
 31. R. Holzinger,  R. Guti\'errez-J\'auregui, T. H\"onigl-Decrinis, G. Kirchmair, A. Asenjo-Garcia, and H. Ritsch, Control of Localized Single- and Many-Body Dark States in Waveguide QED, Phys. Rev. Lett. {\bf 129} 253601 (2022).\\
32. J. Dubois, U. Saalmann, and J.M. Rost, Symmetry-induced decoherence-free subspaces,
Phys. Rev. Research {\bf 5}, L012003 (2023).\\
33. O. Rubies-Bigorda, S.J. Masson, S.F. Yelin, and A. Asenjo-Garcia,
Deterministic generation of photonic entangled states using decoherence-free
subspaces, arXiv:2410.03325 (2024).\\
34.  W. Chen, G.D. Lin, and H.-H. Jen,
Excitation transfer and many-body dark states in WQED, arXiv:2504.12677 (2025).\\
35. J. Skaar, J.C. Garc\'{\i}a Escart\'{\i}n, and H. Landro, Quantum mechanical description of linear optics, Am. J. Phys. {\bf 72}, 1385 (2004).\\
36. S. Longhi, Quantum interference and exceptional points, Opt. Lett. {\bf 43}, 5371 (2018).\\
37. S. Longhi, Quantum statistical signature of PT symmetry breaking, Opt. Lett. {\bf 45}, 1591 (2020).\\
38. S. Longhi, Bosonic Mpemba effect with non-classical states of light , APL Quantum {\bf 1}, 046110  (2025).\\
39. A. S. Sheremet, M. I. Petrov, I. V. Iorsh, A. V. Poshakinskiy,
and A. N. Poddubny, Waveguide quantum electrodynamics:
Collective radiance and photon-photon correlations, Rev.
Mod. Phys. {\bf 95}, 015002 (2023).\\


 \end{document}